\documentclass[aps,longbibliography,showpacs,twocolumn,superscriptaddress,amsmath,amssymb,verbatim]{revtex4-1}

\usepackage[thinlines]{easytable}
\usepackage{graphicx}
\usepackage{lmodern}
\usepackage{epstopdf}
\usepackage{dcolumn}
\usepackage{bm}
\usepackage{subfigure}
\usepackage{amsmath} 
\usepackage[pdftex,colorlinks=true,citecolor=blue,linkcolor=blue,urlcolor=blue,bookmarks=true]{hyperref}

\usepackage{color}
\usepackage{textcomp}
\definecolor{red}{rgb}{1,0,0}

\definecolor{blue}{rgb}{0,0,1}

\definecolor{green}{rgb}{0,1,0}

\begin{document}
	\preprint{APS}
\title{ Development of short and long-range magnetic order in the double perovskite based frustrated triangular lattice antiferromagnet  Ba$_{2}$MnTeO$_{6}$}

\author{J. Khatua}
\affiliation{Department of Physics, Indian Institute of Technology Madras, Chennai 600036, India}
\author{T. Arh}
\affiliation{Jo\v{z}ef Stefan Institute, Jamova c. 39, 1000 Ljubljana, Slovenia}
\affiliation{Faculty of Mathematics and Physics, University of Ljubljana, Jadranska u. 19, 1000 Ljubljana, Slovenia}
\author{Shashi B. Mishra}
\affiliation{Condensed Matter Theory and Computational Lab, Department of Physics, Indian Institute of Technology Madras, Chennai 600036, India}
\author{H. Luetkens}
\affiliation{Laboratory for Muon Spin Spectroscopy, Paul Scherrer Institute, Villigen, PSI, Switzerland}
\author{A. Zorko}
\email{andrej.zorko@ijs.si}
\affiliation{Jo\v{z}ef Stefan Institute, Jamova c. 39, 1000 Ljubljana, Slovenia}
\affiliation{Faculty of Mathematics and Physics, University of Ljubljana, Jadranska u. 19, 1000 Ljubljana, Slovenia}
\author{B. Sana}
\affiliation{Department of Physics, Indian Institute of Technology Madras, Chennai 600036, India}
\author{M. S. Ramachandra Rao}
\affiliation{Department of Physics, Nano Functional Materials Technology Centre and Materials Science Research Centre, Indian Institute of Technology Madras, Chennai-600036, India}
\author{ B. R. K. Nanda}
\affiliation{Condensed Matter Theory and Computational Lab, Department of Physics, Indian Institute of Technology Madras, Chennai 600036, India}
\author{P. Khuntia}
\email[]{pkhuntia@iitm.ac.in}
\affiliation{Department of Physics, Indian Institute of Technology Madras, Chennai 600036, India}

\date{\today}

\begin{abstract}
 Oxide double perovskites wherein octahedra formed by both 3$d$ elements and $sp$-based heavy elements give rise to unconventional magnetic ordering and correlated quantum phenomena crucial for futuristic applications. Here, by carrying out experimental and first principles investigations, we present the electronic structure and magnetic phases of Ba$_{2}$MnTeO$_{6}$, where Mn$^{2+}$ ions with \textit{S} = 5/2 spins constitute a perfect triangular lattice. The magnetic susceptibility reveals a large Curie-Weiss temperature $\theta_{CW} $ = $-$152 K suggesting the presence of strong antiferromagnetic interactions between Mn$^{2+}$ moments in the spin lattice. A phase transition at 20 K is revealed by magnetic susceptibility and specific heat which is attributed to the presence of a sizeable inter-plane interactions. Below the transition temperature, the specific heat data show antiferromagnetic magnon excitations with a gap $\Delta/k_{B}$ $\approx$ 1.4 K. Furthermore, muon spin-relaxation reveals the presence of static internal fields in the ordered state and provides strong evidence of short-range spin correlations for \textit{T} $>$ $T_{N}$. The DFT+\textit{U} calculations and spin-dimer analysis infer that Heisenberg interactions govern the inter and intra-layer spin-frustrations in this perovskite. The inter and intra-layer exchange interactions are of comparable strengths ($J_{1}$ = 4.6 K , $J_{2}$ = 0.92 $J_{1}$). However, a weak third nearest-neighbour ferromagnetic inter-layer interaction exists ($J_{3}$ = $-$ 0.04 $J_{1}$) due to double-exchange interaction via the linear path Mn-O-Te-O-Mn. The combined effect of $J_{2}$ and $J_{3}$ interactions stabilizes a three dimensional (3\textit{D}) long-range magnetic ordering in this compound.
\end{abstract}

\maketitle
\section{Introduction}
Incompatibility of magnetic interactions with a spin lattice leads to  spin frustration and strong quantum
fluctuations yielding  novel states in quantum materials \cite{ANDERSON1973153,Balents2010,Khuntia2020,PhysRevB.96.094432}. Triangular lattice antiferromagnet offers the simplest prototype to realize frustration induced ground states. In this antiferromagnets, the complex interplay between several degrees of freedom and  strong quantum fluctuations lead to various types of intriguing physical phenomena such as quantum spin liquids \cite{PhysRevB.86.140405,PhysRevB.93.140408}, field induced magnetization plateaus \cite{PhysRevLett.102.137201,PhysRevLett.109.267206,PhysRevB.92.180411}, non-colinear 120$^\circ$ magnetic ordered states \cite{PhysRevB.90.224402}, spin-driven ferroelectricity \cite{PhysRevLett.101.067204,PhysRevLett.109.257205}, etc. Experimentally it has been shown that  the triangular lattice with low spin is ideal to realize exotic spin-liquid ground state due  to enhanced zero-point spin fluctuations as observed, e.g., in \textit{k}-(BEDT-TTF)$_{2}$Cu$_{2}$(CN)$_{3}$ \cite{PhysRevLett.91.107001,PhysRevLett.95.177001}, EtMe$_{3}$Sb[Pd(dmit)$_{2}$]$_{2}$ \cite{doi:10.7566/JPSJ.83.034714,PhysRevB.88.075139}
and 1T-TaS$_2$ \cite{klanjvsek2017high}. On the other hand, for classical Heisenberg spins on triangular lattices, inter-planar interaction,  next-nearest-neighbor interaction and anisotropic interaction suppress  low dimensional magnetism resulting in 3\textit{D} magnetic ordered state as observed in  various compounds, e.g.,
RbFe(MoO$_{4}$)$_{2}$ (Fe$^{3+}$, \textit{S} = 5/2) \cite{PhysRevLett.98.267205}, NaBa$_{2}$Mn$_{3}$F$_{11}$ (Mn$^{2+}$, \textit{S} = 5/2) \cite{f11},  
LiCrO$_{2}$ 
( Cr$^{3+}$, \textit{S} = 3/2) \cite{Kadowaki_1995}, Rb$_{4}$Mn(MoO$_{4}$)$_{3}$ (Mn$^{2+}$, \textit{S} = 5/2) \cite{Ishii_2011,doi:10.1143/JPSJ.80.064705}, Ba$_{8}$MnNb$_{6}$O$_{24}$ (Mn$^{2+}$, \textit{S} = 5/2) \cite{PhysRevMaterials.3.054412}, etc. In such classical spin systems strong thermal fluctuations lead to  magnetically ordered ground state via ``order by disorder'' mechanism \cite{PhysRevB.46.11137}.
The physics of frustrated triangular lattice is rich and diverse \cite{ANDERSON1973153,PhysRevLett.112.127203,Starykh_2015}, however, in many cases anti-site disorder, anisotropy and inter-plane interactions put a strong constraint on the  ground state spin dynamics and challenge theoretical paradigms \cite{PhysRevLett.109.117203,PhysRevB.84.245108,PhysRevLett.119.157201}. The current challenge is to explore disorder-free frustrated triangular lattice antiferromagnets with exactly solvable ground state  potential to host exotic magnetism and spin dynamics.\\
In this context, \textit{B} site ordered double perovskites of general formula \textit{A}$_{2}$\textit{B}\textit{B\'}O$_{6}$, where \textit{A} represents a divalent cations, \textit{B} is a 3\textit{d} transition metal ions and \textit{B\'} = Te$^{6+}$, Mo$^{6+}$ or W$^{6+}$ offer an alternate route to realization of novel magnetism and spin dynamics as a result of intricate interplay between spin, lattice and charge degrees of freedom.
It has been observed that many unconventional magnetic ground states are governed by planar structure of \textit{B}-site ions. For example, Ba$_{2}$CoTeO$_{6}$ is a unique case of \textit{B}-site ordered double perovskite, where Co$^{2+}$ (\textit{S} = 1/2) ions form two (triangular and honeycomb) subsystems. The spins on the triangular lattice behave as  Heisenberg spins, while the  spins on the honeycomb lattice show Ising like antiferromagnetic interactions \cite{PhysRevB.93.094420,PhysRevB.96.064419}. Electron-spin resonance (ESR) and magnetization measurements show that applied magnetic field perpendicular to the easy-axis induces magnetization plateaus for both  sub-lattices due to strong quantum effects of \textit{S} = 1/2 spins \cite{PhysRevB.93.094420,PhysRevB.96.064419,B927498G}. Another  interesting example is Sr$_{2}$CuTeO$_{6}$, a quasi-two dimensional Heisenberg antiferromagnet, where Cu$^{2+}$ (\textit{S} = 1/2) ions form a planar square lattice and develop N\'eel order below 29 K \cite{PhysRevB.93.054426}. Recently, magnetic susceptibility, specific heat  and $\mu$SR studies on Sr$_{2}$CuTe$_{1-x}$W$_{x}$O$_{6}$ demonstrated a quantum disordered ground state for \textit{x} = 0.5 \cite{PhysRevB.98.054422}. This is a promising candidate to tune electron correlation by  quenched disorder in the \textit{J}$_{1}$-\textit{J}$_{2}$ Heisenberg model on a square lattice.\\
Recently, a  new \textit{B} site ordered double perovskite Ba$_{2}$MnTeO$_{6}$ (henceforth BMTO), where Mn$^{2+}$ ions with spin \textit{S} = 5/2 constitute a perfect  spin lattice without anti-site disorder, has been reported  \cite{Mustonen2020,PhysRevB.102.094413,bamn}.  While  one of the study proposed a cubic space group (Fm$\bar{3}$m), the other  suggested a trigonal space group R$\bar{3}$m for describing the 
structure of BMTO \cite{Mustonen2020,PhysRevB.102.094413}. The same trigonal space group has been proposed for the BMTO in an earlier study\cite{bamn}.  The high-temperature magnetic susceptibility data follow the Curie-Weiss law with large  Curie-Weiss temperature, which  suggests the
presence of strong antiferromagnetic interaction between
Mn$^{2+}$ spins. An anomaly is observed in the magnetic susceptibility and specific heat data at $T_{N}$ = 20 K, which
is an indication of a symmetry breaking phase transition
in BMTO. However, a clear picture of the crystal structure and exchange interactions of BMTO is missing. Also, the presence of static internal fields, the development of the order parameter and spin dynamics above and below the N\'eel temperature of this novel antiferromagnet have not been yet explored. \\
In this work, we report our results based on XRD, magnetization, specific heat, and muon spin relaxation ($\mu$SR) studies as well as density functional theory calculations in order to shed new insight into the crystal structure, magnetism, anisotropy and spin correlations in this novel frustrated triangular lattice antiferromagnet. We have found that the  trigonal and cubic structures can both index the observed XRD peaks in BMTO, as also concluded previously \cite{ PhysRevB.102.094413}. All structural data thus indicate that the two structures are only marginally different, so that the exact structure likely has no significant effect on magnetism and spin dynamics. Indeed, it is suggested that the  two structures are very close  if one converts one space group to the other. As the trigonal space group offers an additional degree of freedom for the positions of Ba and O sites along the \textit{c}-axis, this suggests that the trigonal space group may be advantageous over the cubic one as reported recently \cite{PhysRevB.93.024412,PhysRevB.102.094413,B927498G,VASALA20151}. The N\'eel ordering at $T_N=20$\,K is confirmed by local-probe $\mu$SR measurements revealing the appearance of static internal fields below $T_{N}$. These measurements show that the whole sample enters a long-range magnetically ordered state below this temperature, while short-range correlations are observed all the way up to 35 K. Moreover, $\mu$SR reveals critical slowing-down of spin dynamics at $T_{N}$ and its persistence to the lowest temperatures, as well as  
tracks the development of the order parameter.  Furthermore, we find that a broad maximum at 10 K in the magnetic specific heat data can be associated with the magnon gap  due to magnetic anisotropy. Namely, the magnetic specific heat data below $T_{N}$ reveals magnon excitations with a gap of $\sim$ 1.4 K. The origin of magnetic ordering in BMTO is studied within the frame work of density functional theory (DFT) calculations for the trigonal space group.  Our  DFT results reveal that the Mn$^{2+}$ spins favor a dominant Heisenberg antiferromagnetic ordering consistent with the experimental results. Our calculations using the DFT +\textit{U} formalism yield intra-layer exchange energy $J_{1}$ = 4.6 K and a comparable inter-layer exchange coupling $J_{2}$ = 0.92 $J_{1}$. In addition, a weaker ferromagnetic inter-layer interaction exists with third nearest neighbor ($J_{3}$ = $-$  0.04 $J_{1}$) due to double-exchange interaction via the linear path Mn-O-Te-O-Mn. Though the strength of this indirect interaction ($J_{3}$) is one order of magnitude smaller than the leading AFM interaction, the combined effect of $J_{2}$ and $J_{3}$ contributes towards stabilizing the long-range magnetic order in this frustrated magnet.
\section{EXPERIMENTAL DETAILS}  Polycrystalline samples of BMTO were prepared  by a conventional solid state  method. Prior to  use, we  preheated  BaCO$_{3}$ (Alfa Aesar, 99.997 \text{\%}), MnO$_{2}$ (Alfa Aesar, 99.996 \text{\%})   and TeO$_{2}$ (Alfa Aesar, 99.9995 \text{\%}) to remove any moisture.  The appropriate stoichiometric mixtures were  pelletized and sintered at 1200$^{\circ}$C for 30 hours with several intermittent grindings.  The phase purity was confirmed by the Rietveld refinement of XRD  taken employing  a smartLAB  Rigaku X-ray diffractometer  with Cu $\mathcal{K\alpha}$ radiation ($\lambda $ = 1.54 {\AA}).
Magnetization measurements were carried out using a Quantum Design, SQUID VSM in the temperature range 5 K $\leq \textit{T}\leq $ 340 K under magnetic fields 0 T $\leq \mu_{0}H\leq$ 7 T. Specific heat measurements were performed on a Quantum Design Physical Properties Measurement System (QD, PPMS) by thermal relaxation method, in the temperature range 2 K $\leq \textit{T}\leq $ 240 K.  $\mu$SR measurements were performed using the GPS spectrometer at the Paul Scherrer Institute, Villigen, Switzerland, on a 1-gm powder sample in the temperature range 1.6 K $\leq$ \textit{T} $\leq$ 50 K. The sample was put on a ``fork" sample holder and the veto mode was employed, which ensured minimal background signal. The transverse muon-polarization was used in zero applied field (ZF) and in a weak transverse field (TF) of 5 mT.
 \section{RESULTS}
   \begin{table}
  	\vspace*{0.5 cm}
  	\caption{\label{tab:table1}Rietveld refinement results at room temperature for BMTO with space group $R\bar{3}m$ and unit cell parameters  \textit{a} = \textit{b} = 5.814 {\AA}, \textit{ c} = 14.243 {\AA}
  		and $\alpha $ = 90$^{\circ}$, $\beta$ = 90$^{\circ}$, $\gamma $ = 120$^{\circ}$. The goodness of Rietveld refinement
  		was confirmed by the following factors: $\chi^{2}$ = 4.8; R$_{wp}$ = 6
  		\text{\%}; R$_{exp}$= 2.72 \text{\%} and 
  		R$_{p}$ = 4 \text{\%}.  }
  	\vspace{5 mm}
  	\begin{tabular}{c c c c c  c c} 
  		\hline \hline
  		Atom & Wyckoff position & x & y & z& Occ.\\
  		\hline 
  		Mn & 3\textit{a} & 0 & 0 & 0 & 1 \\
  		Te & 3\textit{b} & 0 & 0 & 0.5 & 1 \\
  		Ba & 6\textit{c} & 0& 0 & 0.25 & 1 \\
  		
  		O & 18\textit{h} &0.489 &0.51 &0.245 &1 \\
  		
  		\hline
  	\end{tabular}
  \end{table}
\begin{figure*}
	\centering
	\includegraphics[width=\textwidth]{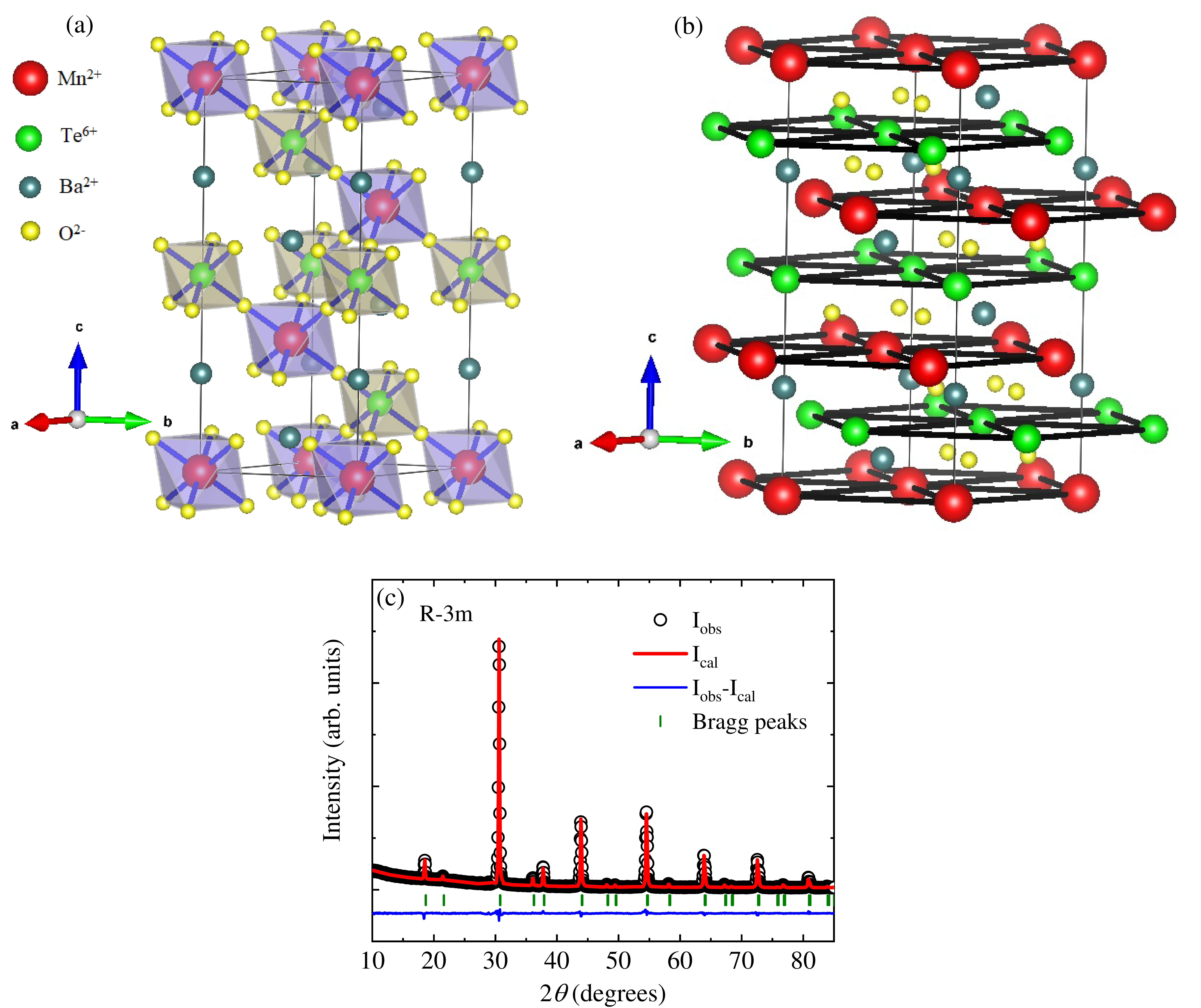}
	\caption{\label{st}(a) The arrangement of MnO$_{6}$ (purple) and TeO$_{6}$ (dark yellow) octahedra in the unit cell of BMTO. (b) Schematic of consecutive triangular layers of Mn$^{2+}$ ions (red) separated by triangular layer of Te$^{6+}$ ions (green) along \textit{c} axis. The \textbf{VESTA} software \cite{doi:10.1107S0021889808012016} was used for visualization of the crystal structure. (c) Rietveld refinement profile of X-ray diffraction data with the solid line (\textit{I}$_{cal}$) through the experimental points (\textit{I}$_{obs}$) calculated for trigonal crystal structure of BMTO. The olive vertical bars indicate the  position of the  Bragg reflections
		and the residual data are denoted by the blue solid line. }	
\end{figure*}
 \subsection{XRD and structural details}
 To check the phase purity, we first measured the
 XRD pattern of polycrystalline BMTO samples. 
 Fig.~\ref{st} (c) depicts the powder XRD pattern at room
 temperature. Rietveld refinement of XRD data using GSAS
 software \cite{doi:10.1107/S0021889801002242} reveals that BMTO crystallizes in
 trigonal crystal structure with the space group R$\bar{3}$m (No.166)
 and gives lattice parameters (Table \ref{tab:table1})  that are consistent with those  previously
 reported \cite{PhysRevB.102.094413,bamn}. Our analysis reveals the
 absence of any site disorder between constituent atoms
 in BMTO. 
\begin{figure*}
		\centering
		\includegraphics[width=\textwidth]{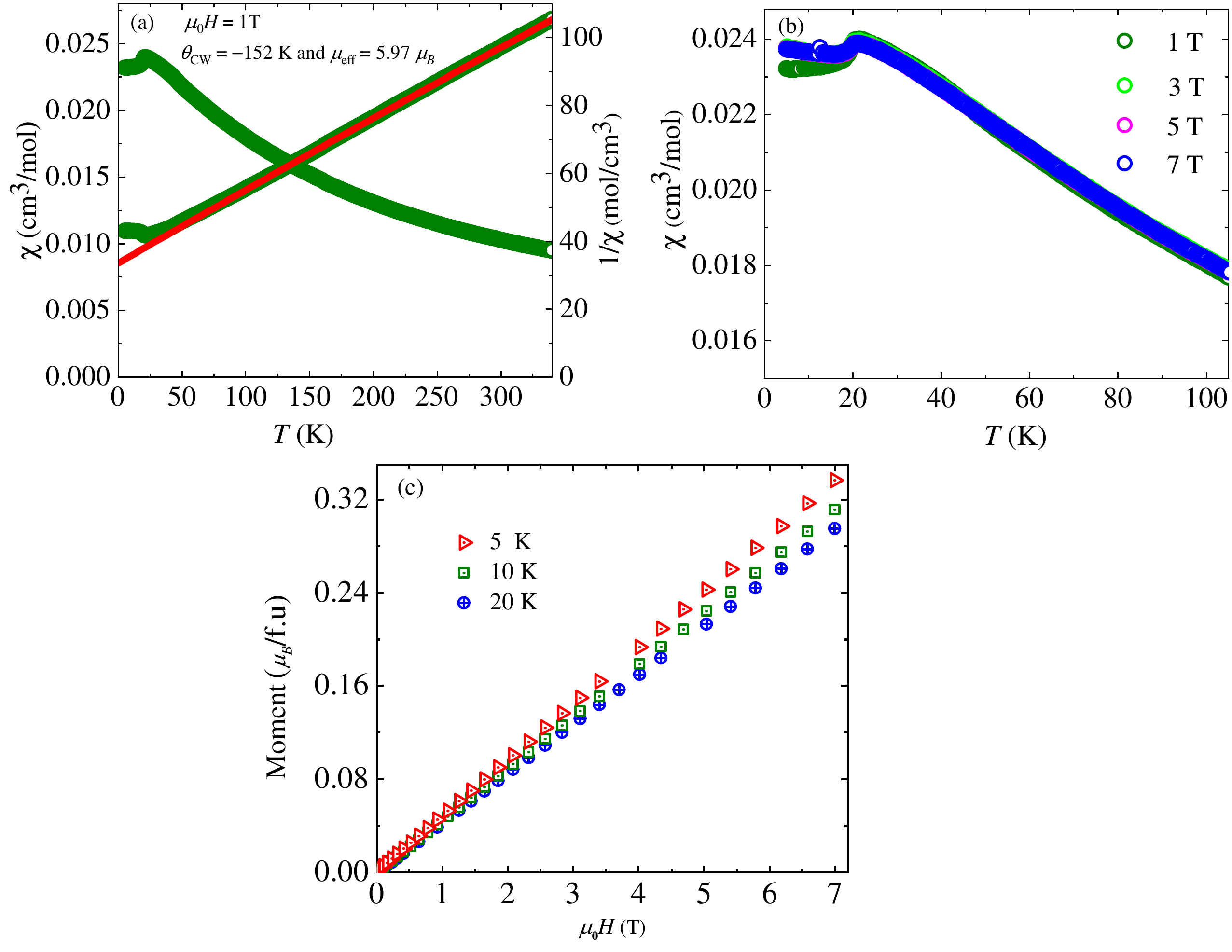}
	\caption{\label{mag}(a) The temperature dependence of dc magnetic susceptibility ($\chi(T)$); left axis) and  inverse magnetic susceptibility (1/$\chi(T)$; right axis ) with the Curie-Weiss fit. (b) The temperature dependence of $\chi(T)$ for different magnetic fields in the temperature range 5 K $\leq$ \textit{T} $\leq$ 100 K. (c) Magnetization vs. field of BMTO at several temperatures. }	
\end{figure*} 
 In the BMTO crystal structure shown in Fig.~\ref{st} (a) Mn$^{2+}$ (3\textit{d}$^{5}$; \textit{S} = 5/2) and Te$^{6+}$  (4\textit{d}$^{10}$; \textit{S} = 0) ions  form MnO$_{6}$ and   TeO$_{6}$ octahedra with nearest-neighbor oxygen ions, respectively. The Mn-O bond length within the MnO$_{6}$ octahedra  is 2.179 {\AA} and the Te-O bond length within the TeO$_{6}$ is 1.932 {\AA}. In the \textit{a}-\textit{b} plane, 
the nearest-neighbor (5.81 {\AA})  Mn$^{2+}$ ions form  equilateral triangular planes stacked along \textit{c} axis (see Fig.~\ref{st} (b)). The consecutive Mn$^{2+}$  triangular planes are separated by non-magnetic triangular planes of Te$^{6+}$ (Fig.~\ref{st} (b)). The adjacent inter-layer Mn-Mn distance is 5.81 {\AA}. The 
nearest-neighbor  Mn$^{2+}$
 ions of MnO$_{6}$ octahedra are connected via TeO$_{6}$ octahedra.
  The inter-planar Mn$^{2+}$ ions are connected through the linear
 path Mn$^{2+}$(1)-O$^{2-}$-Te$^{6+}$-O$^{2-}$-Mn$^{2+}$(2), 
 here Mn$^{2+}$(1)
and Mn$^{2+}$(2) refer to Mn$^{2+}$ ions in two adjacent planes.  We found a 
similarity between the structure of BMTO with rare-earth based spin-liquid 
 candidate YbMgGaO$_{4}$, though the latter is composed of 4\textit{f} magnetic ions \cite{Li2015}. YbMgGaO$_{4}$ crystallizes in the same space group $R\bar{3}m$ with  lattice parameters \textit{a} = \textit{b} = 3.41 {\AA} and \textit{c} = 25.14 {\AA} \cite{Li2015}. In YbMgGaO$_{4}$, a single crystallographic site (3a) of Yb atoms and atomic coordinate (0, 0, 0) matches with coordinate of Mn$^{2+}$ in BMTO. Although both  systems belong to the same  crystal class,  the spin orbit coupling plays an important role to host an exotic ground state in YbMgGaO$_{4}$ \cite{PhysRevLett.117.097201}, whereas in BMTO the inherent physics of high-spin-state of the magnetic ions is expected to be different due to much smaller spin-orbit coupling and the presence of finite inter-plane interactions.
\subsection{\label{sec: Magnetic susceptibility }Magnetic susceptibility }
Fig.~\ref{mag} (a) depicts the temperature dependence  of the magnetic susceptibility ($\chi(T)$) of BMTO in a magnetic field $\mu_{0}H= $ 1 T. In order to estimate the effective magnetic moment $\mu_{eff}$ and the Curie-Weiss temperature ($\theta_{CW}$), the inverse magnetic susceptibility,  1/$\chi(T)$  was fitted (see right y-axis of Fig.~\ref{mag} (a)) with the Curie-Weiss model
\begin{equation}
\chi=\chi_{0}+\frac{C}{T-\theta_{CW}},
\end{equation}
  where $\chi_{0}$ is the temperature independent contribution due to core diamagnetism and van Vleck paramagnetism, \textit{C} is the Curie-constant and $\theta_{ CW}$ is the Curie-Weiss temperature. The Curie-Weiss fitting in the high-temperature range 150 K $\leq $ \textit{T}  $\leq$ 340 K 
 yields \textit{C} = 4.45 cm$^{3}$ K/mol, $\chi_{0}$ = 4.5 $\times$ 10$^{-5}$ cm$^{3}$/mol and 
 $\theta_{ CW}$ = $-$ 152 K.
  The relatively large and negative value of $\theta_{CW}$
  reveals  the presence of strong antiferromagnetic exchange interaction between Mn$^{2+}$ spins. The calculated  effective moment $\mu_{eff} = \sqrt{8C}= 5.97 \mu_{B}$ per Mn atom, which is very  close to the expected moment  $\mu_{eff} = g \sqrt{S(S+1)} \mu_{B} = 5.92  \mu_{B}$ for the high-spin state (\textit{S} = 5/2) of Mn$^{2+}$  assuming the $g$ factor $g = 2$ \cite{PhysRevB.90.024431}. The high spin state is further confirmed from the DFT calculations.
  The corresponding effective moment  gives the Land\'e g factor
  $g = 2.018$, a similar $g$ value was  also
 determined in the triangular lattice Ba$_{3}$MnSb$_{2}$O$_{9}$ by ESR \cite{TIAN201410}. 
 With decreasing 
temperature the $\chi(T)$ data start deviating 
from the Curie-Weiss law and show an anomaly at 20 K which suggests  an antiferromagnetic long-range  order  at this temperature.
 Similar behavior was also seen in several other frustrated triangular lattice systems \cite{PhysRevB.78.104427}.  Indeed, other compounds in this series of double perovskites also show a common feature of long-range antiferromagnetic ordering around 20 K including Sr$_{2}$CuTe$O_{6}$ and Pb$_{2}$MnTeO$_{6}$ \cite{PhysRevB.93.054426,Retuerto2016}. The strength of frustration in the present antiferromagnet is
quantified by the frustration parameter \textit{f} = $\frac{|\theta_{CW}|}{T_{N}} \approx$ 7, which suggests the existence of
moderate frustration in the host magnetic lattice. As shown in Fig.~\ref{mag} (b), $\chi(T)$ data for all fields up to 7 T are very similar in magnitude and we observed no shift in anomaly with the applied field up to 7 T. Absence of any hysteresis in magnetization curve at 5 K (Fig.~\ref{mag} (c)) excludes any ferromagnetic component, either being intrinsic or due to a tiny amount of  impurity phase of Mn$_{3}$O$_{4}$ \cite{doi:10.1002/ejic.200500880}. The reduced magnetic moment compared to saturation moment 5.92 $\mu_{B}$ (Mn$^{2+}$, \textit{S} = 5/2) at 7 T
\begin{figure*}
		\centering
		\includegraphics[width=\textwidth]{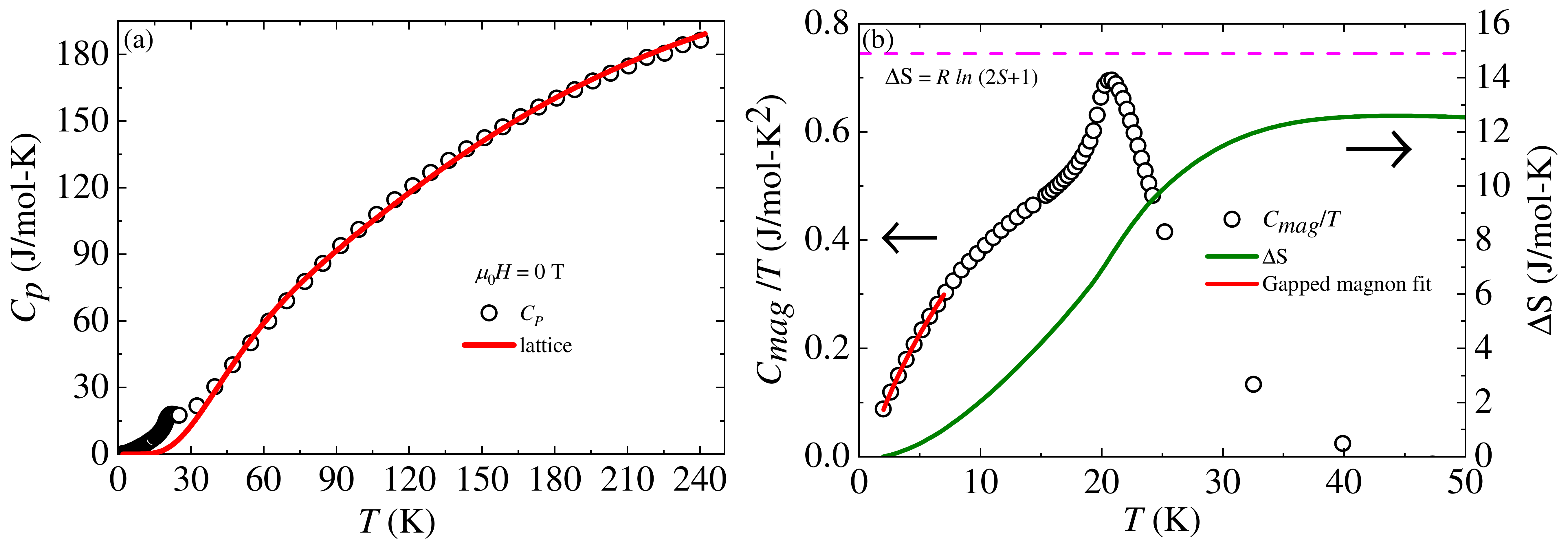}
	\caption{\label{hc}(a) The temperature dependence of specific heat $C_{p}(T)$ of polycrystalline samples of BMTO in the temperature range 2 K $\leq$ \textit{T}  $\leq$ 250 K at zero applied field. The solid line is the fit of $C_{p}(T)$ data to Debye $+$ Einstein model accounting for the lattice  contribution to specific heat. (b) The temperature dependence of magnetic specific heat ($C_{mag}(T)/T)$ of BMTO in the temperature range 2 K $\leq$ \textit{T} $\leq$ 50 K (left y-axis) and the calculated entropy change  with temperature (right y-axis). } 
\end{figure*}
 is consistent with the presence of strong antiferromagnetic exchange interactions between Mn$^{2+}$ spins.
 \subsection{Specific heat}
In order to provide further evidence of long-range magnetic order, we have measured the temperature dependence of specific heat ($C_{p}(T)$) of BMTO  in zero field in the temperature range 2 K $\leq$  \textit{T} $\leq$ 250 K. A lambda-like anomaly appears at $T_{N}$ = 20 K, which is the same temperature at which we observed an anomaly in $\chi(T)$. This  confirms the occurrence of an antiferromagnetic long-range order in BMTO  at this temperature. The absence of any  anomaly at 42 K, which is the transition temperature of Mn$_{3}$O$_{4}$, indicates BMTO is free from minor impurity phase of Mn$_{3}$O$_{4}$ \cite{ROBIE1985165}. An estimate of the associated magnetic contribution to the specific heat data of BMTO is obtained after subtraction of lattice contribution  from the total specific heat data, i.e., $C_{mag}(T)=C_{p}(T)-C_{latt}(T)$, where $C_{mag}(T)$ and $C_{latt}(T)$ are the magnetic and lattice specific heat, respectively. In the absence of a suitable non-magnetic analog of BMTO, we model the lattice contribution as \cite{kittel1976introduction}
\begin{equation*}
C_{latt}(T)=C_{D}[9k_{B} \left(\frac{T}{\theta_{D}}\right)^{3}\int_{0}^{\theta_{D}/T}\frac{x^{4}e^{x}}{(e^{x}-1)^{2}}dx]
\end{equation*}
\begin{equation}\label{eqn:debye}
+\sum_{i=1}^{3} C_{E_{i}}[3R\left(\frac{\theta_{i}}{T}\right)^{2}\frac{exp(\frac{\theta_{E_{i}}}{T})}{(exp(\frac{\theta_{E_{i}}}{T})-1)^{2}}], 
\end{equation}
which includes a Debye term and three Einstein terms. In Eq.(\ref{eqn:debye}) $\theta_{D}$ is the Debye temperature, $\theta_{i}$ are the Einstein temperatures of the three modes, \textit{R} and \textit{k}$_{B}$ are the molar gas constant and Boltzmann constant, respectively.   
As depicted in Fig.~\ref{hc} (a), the  experimental data show good agreement with the model for temperatures above 40 K for $\theta_{D}$ = 324 K, $\theta_{E_1}$ = 128 K, $\theta_{E_2}$ = 194 K, and $\theta_{E_3}$ = 645 K. In the fit the coefficients  were fixed in the ratio C$_{D}$:C$_{E_1}$:C$_{E_2}$:C$_{E_3}$ = 1:1:3:5 as in BMTO the number of acoustic and optical modes of lattice vibration has the ratio of 1:9 \cite{PhysRevB.90.035141}. The one  Debye term corresponds to the acoustic mode and three Einstein terms approximate all optical modes.
After subtracting the lattice contribution, the  magnetic contribution to  specific heat $C_{mag}(T)$  is obtained and shown in Fig.~\ref{hc} (b).  There is a clear anomaly in $C_{mag}(T)/T$ at $T_{N}$ $\sim$ 20 K, which suggests that Mn-Mn exchange interaction connectivity in BMTO is essentially 3\textit{D}. Next, we have calculated the entropy change ($\Delta S(T)$) by integrating  $C_{mag}(T)/T$ over the temperature range  from 2 K to 50 K as shown in Fig.~\ref{hc} (b). It is noticed that the rise of entropy change  with increasing temperature saturated to a value of 12.34 J/mol-K at 50 K, which is somewhat lower than the theoretical value of the total entropy 14.9 J/mol.K (\textit{R} $ln(2\textit{S}+1)$) for the high-spin  \textit{S} = 5/2 state of Mn$^{2+}$ ions.  Thus, we recovered 82 \text{\%} of the expected total entropy  and the missing 18 \text{\%} is most likely due to over-estimation of the lattice contribution to total specific heat and thus underestimation of short-range spin correlations  above $T_{N}$. Below the transition temperature, the lattice contribution to the specific heat becomes practically negligible so the measured specific heat is of magnetic origin. At low temperatures up to $T_{N}$, approximately 50 \text{\%} of the entropy is recovered, suggesting that the other 50 \text{\%} is due to short-range spin correlations that develop already above $T_{N}$.

In order to investigate the nature of  magnetic excitations in the ground state, the low temperature ($\sim$$T_{N}$/3) magnetic specific heat data are fitted with a phenomenological model \cite{PhysRevB.78.104406,Hardy_2016,PhysRevB.96.041405}
\begin{equation}
C_{mag}(T) = \alpha T^{n}exp(-\Delta/T),
\label{eqn:ga}
\end{equation}
 where $\alpha$ and \textit{n} are constants and $\Delta$ is the gap between lower band and upper band of closely spaced energy levels.
 A similar empirical formula was employed to describe the gapped magnon excitations of $\alpha$-RuCl$_{3}$ in the ground state \cite{PhysRevB.96.041405,Jana2018}. The fit yields a gap $\Delta/k_{B} \approx $ 1.4 $\pm$ 0.1 K in the magnetic excitation spectrum. The presence of small gap is attributed to an easy axis anisotropy term in the spin Hamiltonian \cite{Hardy_2016,PhysRevB.83.140401}. 
 \subsection{Muon spin  relaxation ($\mu$SR)}
TF $\mu$SR measurements are a very efficient probe of magnetic ordering and spin correlations.
In the absence of static internal magnetic fields of electronic origin, the muon asymmetry precesses in a weak external transverse field $B_{\rm TF}$ with the frequency $\gamma_\mu B_{\rm TF}/(2\pi)$, where $\gamma_\mu=2\pi\times 135.5$\,MHz/T is the muon gyromagnetic ratio. Muons experiencing additional static internal fields, which are in insulators usually in the range between a few tens and a few hundreds of mT \cite{yaouanc2011muon}, oscillate much faster and lead to a strongly damped signal that is observable only at very short times. 
Except from these short times, muon asymmetry follows the general form 
\begin{equation}
A_{TF}(t) = A_0 \cos\left( \gamma_\mu B_{\rm WT} t\right) {\rm e}^{-\lambda_T t} + A_1,
\label{TF}
\end{equation}  
where the amplitude $A_0$ describes the volume fraction of the sample that experiences zero static internal fields and $A_1>0$ arises from the ordered part of the sample with the internal field being parallel to $B_{\rm TF}$.

\begin{figure*}
	\centering 
	\includegraphics[width=\textwidth]{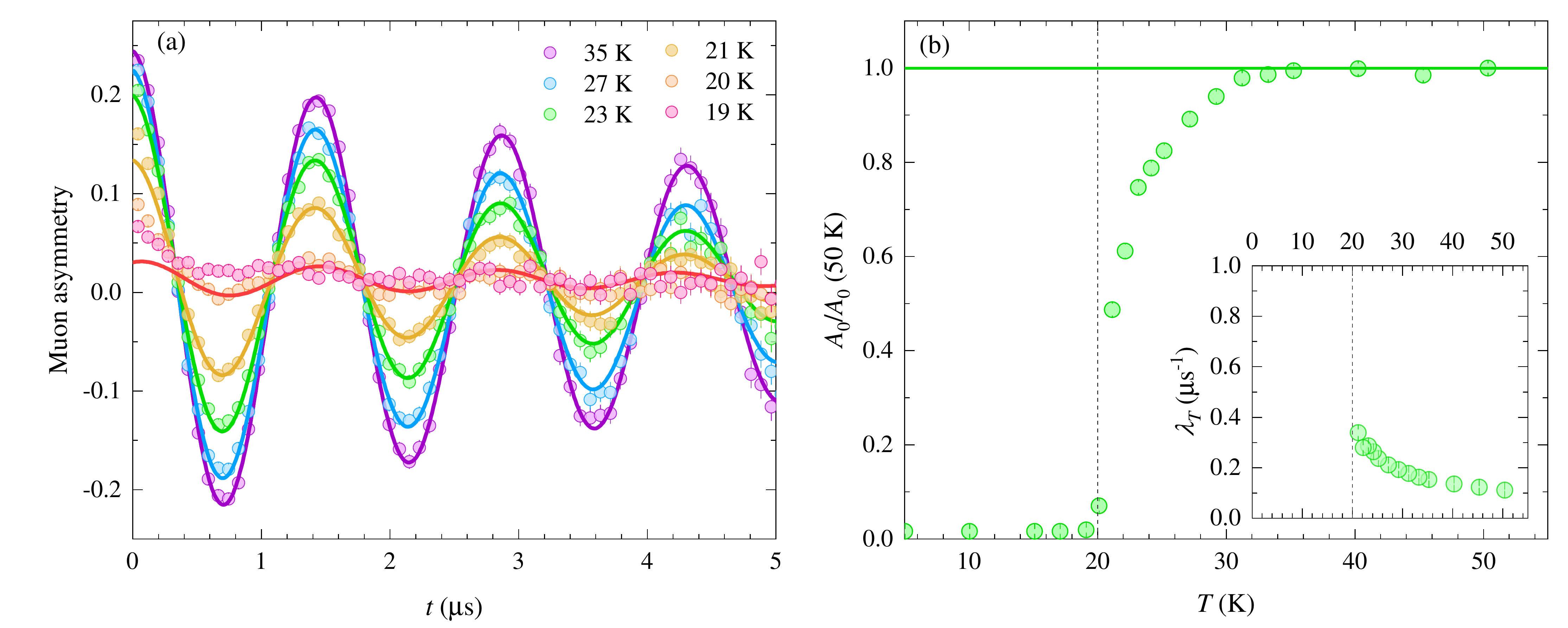}
	\caption{(a) The time dependence of the muon asymmetry in BMTO in a transverse field $B_{\rm TF}=5$\,mT at various temperatures (symbols).
		The solid lines are fits with the model (\ref{TF}) for $t>0.3$\,$\mu$s.
		(b) The relative amplitude of the signal oscillating with the frequency $\gamma_\mu B_{\rm TF}/(2\pi)$ corresponding to the fraction of muons not detecting static internal fields of electronic origin in BMTO.
		The inset shows the temperature dependence of the transverse muon relaxation rate of the oscillating signal. 
		The vertical dashed line  shows the position of $T_N$.	}
	\label{Mu1}
\end{figure*}

\begin{figure*}
	\centering 
	\includegraphics[width=\textwidth]{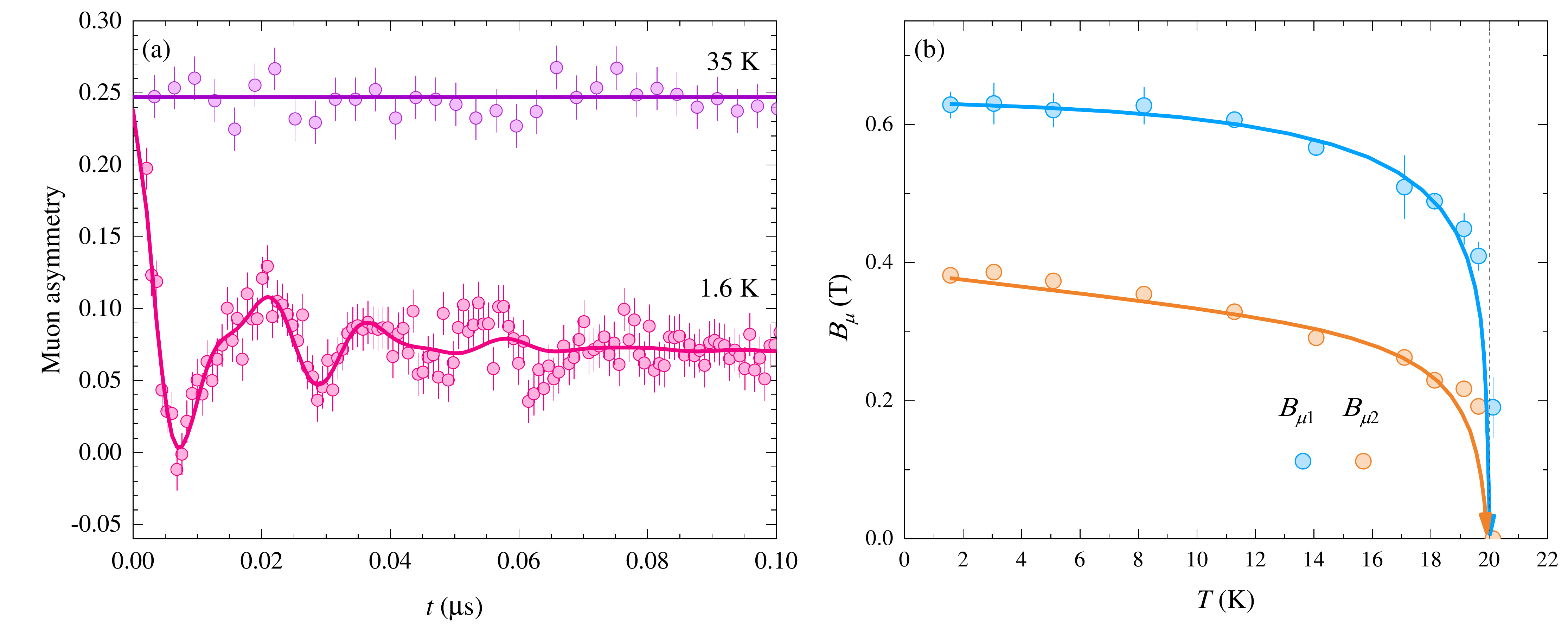}
	\caption{(a) Zero-field muon asymmetry in BMTO at short times above and below $T_N=20$ K (symbols). 
		The solid lines are fits with the model (\ref{order}).
		(b) The temperature dependence of the internal fields at the two muon stopping sites.
		The banded arrows are a guide to the eye and the vertical dashed line shows the position of $T_N$.}
	\label{Mu2}
\end{figure*}

The temperature dependence of the TF $\mu$SR asymmetry in BMTO with corresponding fits of the model (\ref{TF}) is shown in Fig.~\ref{Mu1} (a).
The relative amplitude of the signal oscillating with the frequency $\gamma_\mu B_{\rm TF}/(2\pi)$, i.e., the volume fraction of the muons not experiencing sizeable static internal magnetic fields, starts decreasing from unity already below 35\,K\,$\simeq 2 T_N$ and quickly drops towards zero when the temperature approaches $T_N$ (Fig.~\ref{Mu1} (b)).
$A_0(T)/A_0(50\,{\rm K})<1$ indicates the presence of static internal fields, which we attribute to short-range ordering for $T>T_N$ and long-range ordering for $T<T_N$. 
\begin{figure*}
	\centering 
	\includegraphics[width=\textwidth]{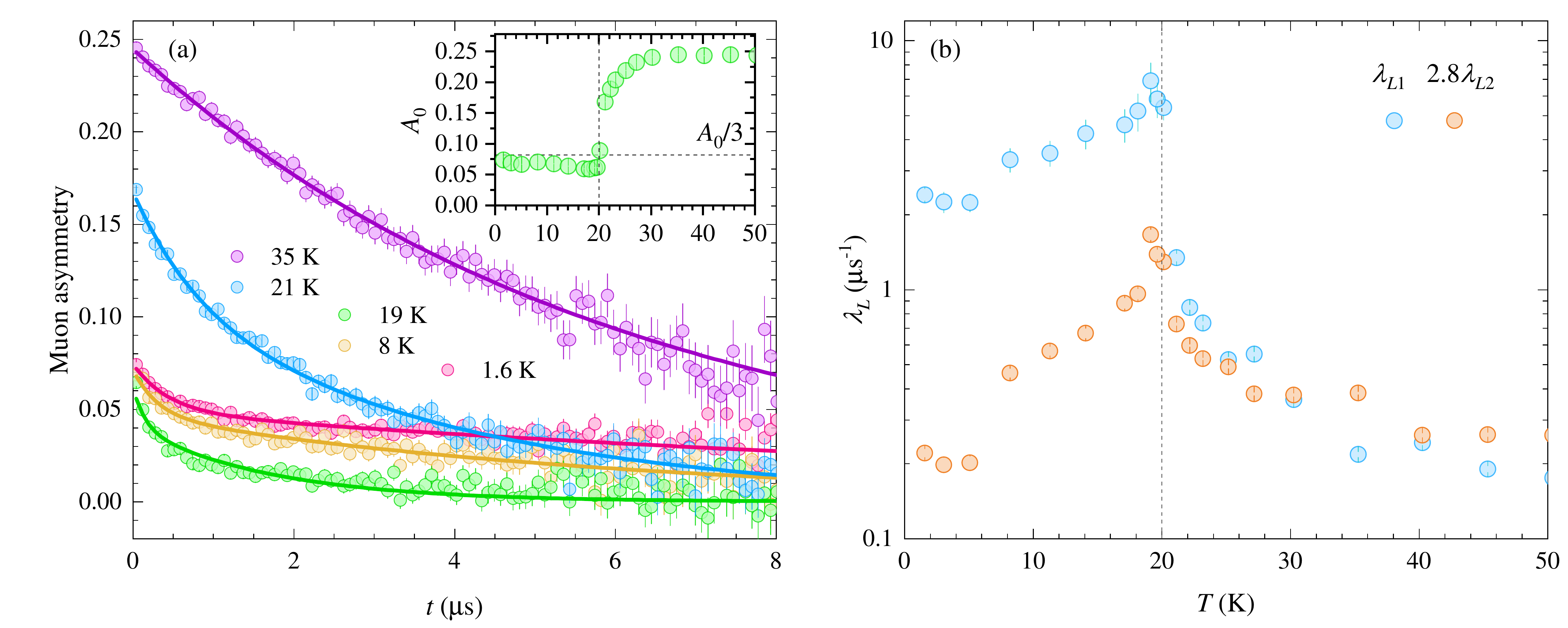}
	\caption{(a) Zero-field muon asymmetry in BMTO at various temperatures (symbols).
		The inset shown the initial asymmetry of the non-oscillating signal.
		The solid lines are fits with the model (\ref{ZF}).
		(b) The temperature dependence of the longitudinal muon spin relaxation rates at the two muon stopping sites in BMTO. 
		The vertical dashed line shows the position of $T_N$.}
	\label{Mu3}
\end{figure*}   
Also in BMTO diffuse neutron scattering originating from the same $Q$ positions as spin waves below $T_N$ is found at temperatures far above $T_N$ \cite{Mustonen2020}, therefore our confirmation of the short-range order nicely complements these results.
We note that the transverse muon spin relaxation rate $\lambda_T$ (inset in Fig.~\ref{Mu1} (b)), which measures the width of the distribution of static fields for the component with no net internal field, increases when temperature approaches $T_N$.\\
Next, we determine the static internal magnetic fields $B_\mu$ in BMTO below $T_N$ more precisely from ZF $\mu$SR measurements, where the frequency of oscillations in muon asymmetry is directly given by these fields, $\nu_\mu=\gamma_\mu B_\mu/(2\pi)$.
Indeed, high-frequency oscillations develop in the muon asymmetry below $T_N$ due to magnetic ordering (Fig.~\,\ref{Mu2} (a)).
The corresponding experimental curves can be fit at short times with a model including two distinct muon stopping sites,
\begin{align}
\nonumber
A_{ZF}^s(t) &= A_0 f \left[\frac{2}{3} \cos\left( \gamma_\mu B_{\mu 1} t\right) {\rm e}^{-\lambda_{T1} t} + \frac{1}{3} \right] \\
&+ A_0 (1-f) \left[\frac{2}{3} \cos\left( \gamma_\mu B_{\mu 2} t\right) {\rm e}^{-\lambda_{T2} t} + \frac{1}{3} \right].
\label{order}
\end{align}  
Here, the constant ``1/3-tail" for each site corresponds to the projection of the initial polarization in a powder sample on the internal magnetic field and, which does not precess, while the oscillating part is due to the perpendicular component \cite{yaouanc2011muon}.
The internal fields at the two muons stopping sites at 1.6\,K amount to $B_{\mu 1}=0.63$\,T and $B_{\mu 2}=0.38$\,T, while large relaxation rates $\lambda_{T 1}\simeq\lambda_{T 2}\sim 60(10)$\,$\mu$s$^{-1}$ indicate relatively broad distributions of internal fields.
The temperature dependence of the average internal fields (Fig.~\,\ref{Mu2} (b)) corresponds to the evolution of the order parameter and the same ratio $B_{\mu 1}/B_{\mu 2}=1.7(1)$ is maintained up to $T_N$. At $T_N$ the static internal fields vanish, contrary to the refined magnetic moment deduced from neutron diffraction, which exhibits a smooth evolution across the transition temperature \cite{Mustonen2020}. 
The fraction of the muon stopping site with the higher internal field value is $f=0.32(5)$  and is temperature independent.
On a longer time scale, the fast oscillations due to static internal magnetic fields below $T_N$ are averaged out, so that only the ``1/3-tail" is seen in ZF muon asymmetry (Fig.~\ref{Mu3} (a)).
This tail exhibits pronounced relaxation even at the lowest temperature of 1.6 K, i.e., well below $T_N$, which is due to the dynamics of the local fields.
In fact, the ZF muon asymmetry on the long time scale can be fit with the same model
\begin{equation}
A_{ZF}^l(t) = A_0 f {\rm e}^{-\lambda_{L 1} t} + A_0 (1-f) {\rm e}^{-\lambda_{L 2} t},
\label{ZF}
\end{equation} 
at all temperatures, where the two terms again correspond to the two muon stopping sites.
The initial asymmetry falls from the high-temperature value of $A_0=0.245$ to about $A_0/3$ (inset in Fig.~\,\ref{Mu3} (a)), as expected.
The decrease of this parameter is due to ordering effects. It  is again observed already below 35\,K and becomes very pronounced in close vicinity of $T_N$, mimicking the change of the amplitude of the TF signal shown in Fig.~\ref{Mu1} (b). 
The longitudinal muon relaxation rates $\lambda_{L 1}$ and $\lambda_{L 2}$ exhibit divergent behavior at $T_N$ (Fig.~\,\ref{Mu3} (b)), which is a typical fingerprint of critical slowing down of spin fluctuations.
Above $T_N$ the ratio of the relaxation rates of the two components scales with the ratio of the squares of the internal fields in the long-range ordered phase, $\lambda_{L 1}/\lambda_{L 2}=(B_{\mu 1}/B_{\mu 2})^2=2.8$ (Fig.~\ref{Mu3} (b)).
As the muon spin relaxation rate is proportional to the square of the fluctuating fields in the fast-fluctuation regime corresponding to the paramagnetic phase \cite{yaouanc2011muon}, this experimental scaling firmly validates our analysis with two muon stopping sites in BMTO.
Below $T_N$ the longitudinal muon spin relaxation is due to collective excitations and the ratio $\lambda_{L 1}/\lambda_{L 2}$ increases to about 10. Importantly, we find that the dynamics of local fields persists down to the lowest temperatures, as observed  in various different frustrated spin systems \cite{PhysRevB.99.214441,PhysRevLett.109.227202,PhysRevB.91.104427}.
 \section{DENSITY FUNCTIONAL STUDIES} 
 \subsection{Computational Methods} 
 To further understand the magnetic interactions in BMTO, DFT calculations have been performed using the plane-wave-pseudopotential approach as implemented in Quantum ESPRESSO \cite{Giannozzi_2017}. The experimentally obtained structure has been considered for the calculations. The ultra-soft pseudopotentials are used to describe the electron-ion interactions \cite{PhysRevB.41.7892}, in which the valence states of Mn include 15 electrons from 3s, 3p, 4s and 3d; Ba includes 10 electrons in 5s, 5p, and 6s; Te includes 10 electrons in 5s and 5p orbitals; and O includes 6 electrons from 2s and 2p shells. The exchange-correlation functional is approximated through PBE-GGA functional \cite{PhysRevLett.77.3865}. The convergence criterion for self-consistent energy is taken to be 10$^{-6}$ Ry. A k-mesh of 4 $\times$ 4 $\times$ 2 is used for the Brillouin zone integration of the supercell of size 2 $\times$ 2 $\times$ 2. The kinetic energy cut-off for the electron wave functions is set at 30 Ry and the augmented charge density cut-off is set to be 300 Ry. We have also performed test calculations with a higher energy cut-off of 40 Ry and charge density cutoff at 400 Ry as well as with a higher k-mesh of 8 $\times$ 8 $\times$ 4. As the results remain the same below the tolerance level, we have used the lower cut-off and lower k-mesh to reduce the computational time. The strong correlation effect is examined through Hubbard\textit{ U} formalism \cite{PhysRevB.71.035105}. The magnetic coupling strengths are evaluated as a function of\textit{ U} in this strongly correlated system.
 \subsection{Electronic Structure}
\begin{figure*}
 	\centering
 	\includegraphics[width=\textwidth]{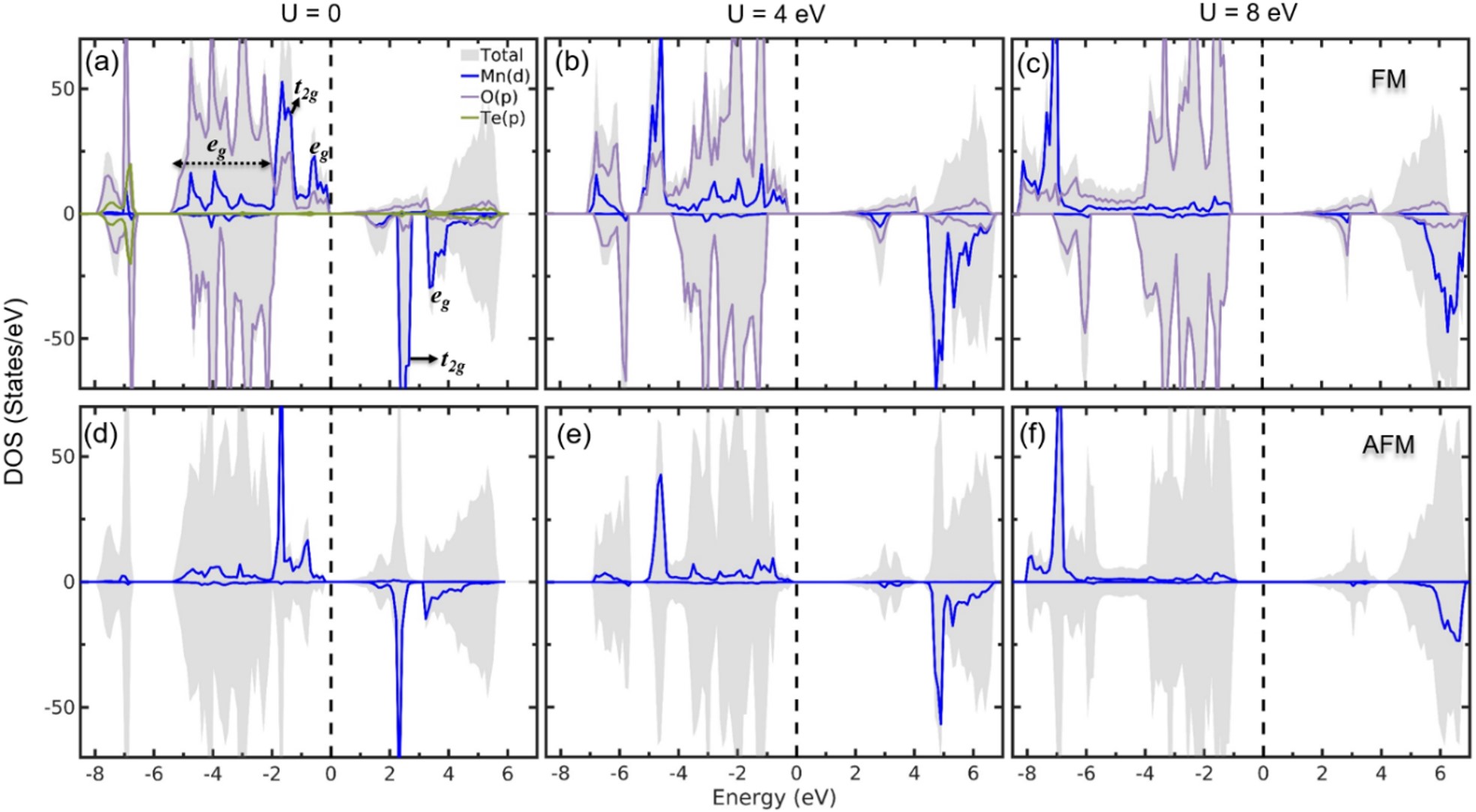}
 	\caption{(a) The total and partial densities of states (DOS) for the ferromagnetic and the energetically most stable AFM configuration (AFM3, see Fig.~\ref{j} (e)) as a function of the Hubbard \textit{U}. For clarity, in the AFM configuration, the Mn-\textit{d} states of only one spin-sublattice is shown. For the opposite spin-sublattice, the corresponding DOS is simply a mirror image. }
 	\label{DOS}
 \end{figure*}

 The crystal structure of BMTO can be described as alternate stacking of layers of TeO$_{6}$ and MnO$_{6}$ octahedra, and the neighboring layers are connected through corner sharing oxygen as shown in Fig.~\ref{st} (a). However, the electronic structure, presented in Fig.~\ref{DOS} through total and partial densities of states (DOS), shows that the Te-\textit{p} state is almost  completely occupied and lie around 7 eV below the Fermi energy ($E_{F}$). Therefore, from the electronic and magnetic structure point of view this compound can be treated as an open spaced structure in the sense that the minimum Mn-Mn separation, both inter-layer and intra-layer, is $\sim$ 5.8 {\AA} which is roughly double than that of the closed packed transition metal perovskites.  Here, the electronic structure of the system is supposed to be nearly a sum of the electronic structure of the individual MnO$_{6}$ octahedra \cite{PARIDA2018133}. To verify this we first examine the DOS within the independent electron approximation (\textit{U} = 0) which are shown in Fig.~\ref{DOS} (a).
  Here, we observe that due to crystal-field splitting, the Mn-\textit{d} states split into triply degenerate \textit{t}$_{2g}$ and doubly degenerate $e_{g}$ states. Due to stronger axial interactions, the $e_{g}$ states have reasonable overlapping with the O-\textit{p} states, which dominate the energy window $-$5 to $-$2 eV with respect to $E_{F}$. Due to 2+ charge state for Mn, the \textit{d}-orbitals are half-occupied and stabilize in a high-spin states (\textit{S} = 5/2) where the \textit{d}-orbitals in the spin minority channel are completely unoccupied. The \textit{S} = 5/2 state is also confirmed from the magnetization measurement discussed earlier.  For such a spin-state, the spin splitting is strong enough to create a band gap even for \textit{U} = 0 \cite{PARIDA2018133}.  With inclusion of strong correlation effect (finite \textit{U}), the Mn-\textit{d} states  are pushed  to lower energies in the valence band and to higher energies in the conduction band to widen the bandgap (see Fig.~\ref{DOS} (a-c)).  The O-\textit{p} states now dominate the valence band near  $E_{F}$ which implies that BMTO is a charge transfer insulator \cite{Jena2016} which favors antiferromagnetic (AFM) ordering. The total and Mn-\textit{d} DOS for the stable AFM ordering (see Fig.~\ref{DOS} (e)) are shown in the lower panel of Fig.~\ref{DOS} (d-f).  As both FM and AFM ordering makes the system insulating, we infer that this is primarily a weakly coupled classical spin system. The strength of the coupling is discussed next. 
  \subsection{Magnetic Interactions in BMTO}
    \begin{figure*}
  	\centering
  	\includegraphics[width=\textwidth]{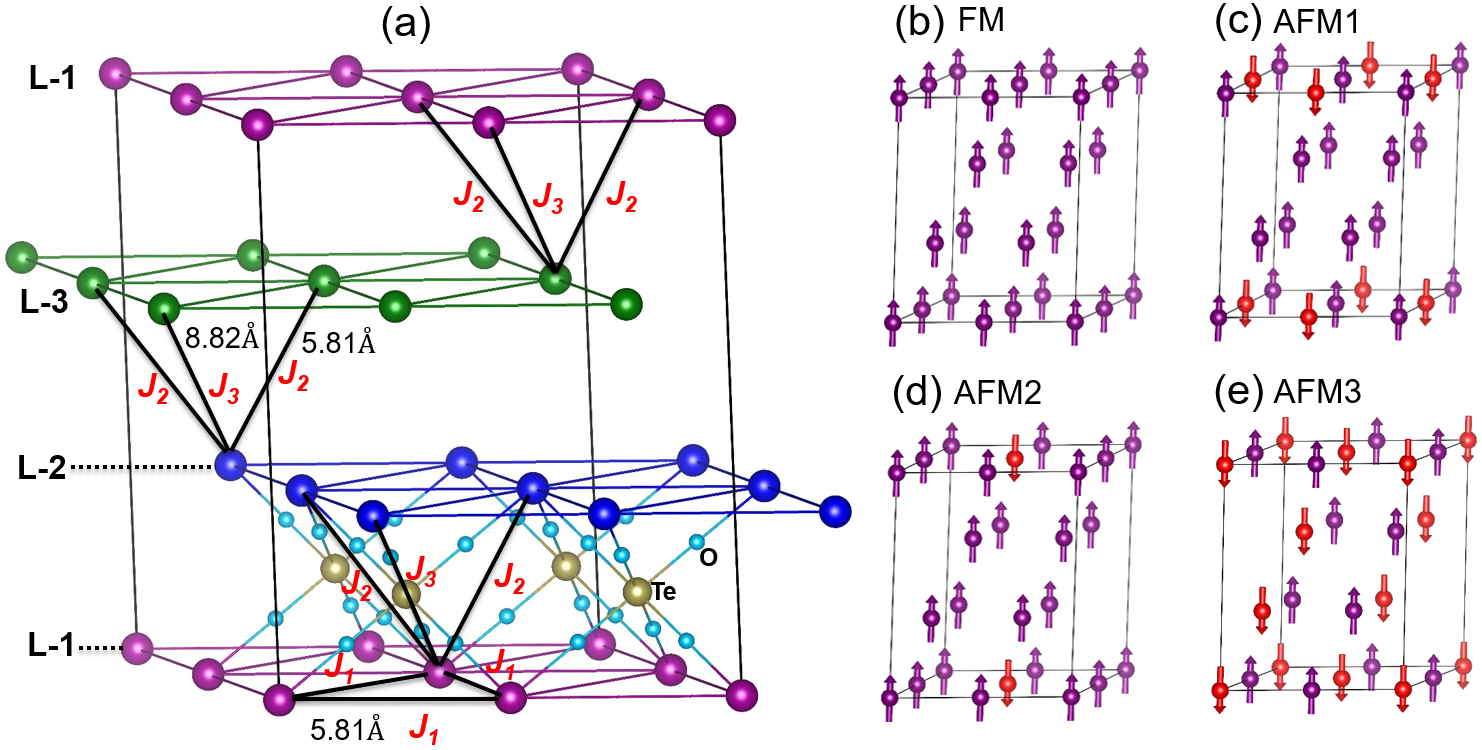}
  	\caption{(a) The dominant exchange interaction paths $J_{1}$, $J_{2}$ and $J_{3}$   for the Mn spins in the layered compound BMTO.  $J_{1}$ represents the in-plane nearest neighbor (5.81 {\AA}) exchange interaction, while $J_{2}$ and $J_{3}$ correspond to out-of-plane nearest and next-nearest interactions with Mn-Mn separations at 5.81 {\AA} and 8.22 {\AA}, respectively. The $J_{1}$ forms a triangular lattice in the plane. To estimate the strengths of the $J_{i}$’s, several total energy calculations are carried out on four spin arrangements, defined as FM, AFM1, AFM2, and AFM3 (b-e). }
  	\label{j}
  \end{figure*}
The experimental results presented in this work imply dominant antiferromagnetic interactions through $\theta_{CW}$. There are three dominant exchange interaction paths ($J_{1}$, $J_{2}$, $J_{3}$) in this compound which demonstrates a hexagonal ABC stacking pattern as shown in Fig.~\ref{j} (a). The $J_{1}$ represents the intra-plane nearest-neighbor interaction for which the Mn-Mn distance is 5.81 {\AA}, whereas $J_{2}$ represents the inter-plane nearest-neighbor interaction ($d_{Mn-Mn}$ = 5.81 {\AA}). The 3rd nearest-neighbor interaction (8.22 {\AA}) is considered by the $J_{3}$ term.  Here, we shall examine the strengths of  these $J_{i}$'s through a spin-dimer analysis using Noodlemann’s broken-symmetry method \cite{doi:10.1063/1.440939,dai2005analysis}. According to this method, the energy difference between the high spin (HS) and low spin (LS) states for a spin dimer is given by
\begin{equation}
E_{HS}-E_{LS} = \frac{1}{2} \left(S_{max}\right)^{2} J,
\label{dimere}
\end{equation}
where \textit{J} is related to the spin-dimer Hamiltonian, $\hat{H}$= \textit{J} $\hat{S}_{1}$ $ \cdot $ $\hat{S}_{2}$, with \textit{ S}$_{max}$ being the maximum
spin of the dimer. As we have a Mn-Mn spin dimer, both sites of the dimer have five  unpaired electrons. Therefore, Eq.(\ref{dimere}) reduces to 
\begin{equation}
E_{HS}-\ E_{LS}=\frac{25}{2}\ J
\end{equation}
where, $E_{HS}$ and $E_{LS}$ can be estimated from the DFT calculations as discussed below.\\
 To evaluate exchange constants in the framework of DFT, one needs to design several possible magnetic configurations, and calculate the total energies. The relative energy differences among them are expressed in terms of $J_{i}$'s leading to a set of linear equations. The magnetic configurations (FM, AFM1, AFM2, and AFM3), considered here are designed on a 2 $\times$ 2 $\times$ 2 supercell as shown in Fig.~\ref{j} (b-e). The total energies of each configuration is estimated with the sum of all exchange paths  which yield the following set of equations:   
\begin{equation*}
\begin{aligned}
  E_{FM} &=& (25/4)(36 J_{1} + 36J_{2} + 36J_{3}) -- (9a)\\   
E_{AFM1} &=& (25/4) (20 J_{1} + 12J_{2} + 12 J_{3}) -- (9b) \\   
 E_{AFM2} &=& (25/4) (24 J_{1} + 24J_{2} + 24 J_{3}) -- (9c) \\ 
E_{AFM3} &=& (25/4) (-12 J_{1} - 12J_{2} + 36 J_{3}) -- (9d)\\ 
\end{aligned}  
\end{equation*}
 \begin{figure}
 	\hspace{-0.5 cm}
	\includegraphics[width=9cm, height= 7 cm]{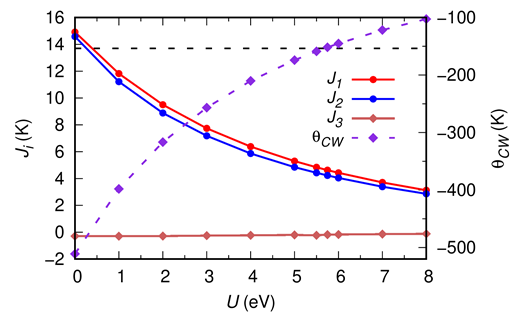}
	\caption{The magnetic exchange-coupling constants of Mn atoms for intra-layer $J_{1}$, and inter-layer $J_{2}$ and $J_{3}$ as well as $\theta_{CW}$ as a function of Hubbard \textit{U}. The experimental $\theta_{CW}$ value ($-$152 K) is shown as dashed line, which matches exactly for \textit{U} = 5.75 eV.}
	\label{plotthe}
\end{figure}
Hence, by solving the above equations, we have estimated the $J_{i}$’s as a function of \textit{U} and plotted  them in Fig.~\ref{plotthe}. While the dominant interactions $J_{1}$ and $J_{2}$ are antiferromagnetic, $J_{3}$ corresponds to a weakly ferromagnetic coupling. This is due to the fact that $J_{3}$ is formed by a linear Mn-O-Te-O-Mn path, where the axial e$_{g}$-O-p-Te-p-O-p-e$_{g}$ covalent interaction is formed leading to a double-exchange ferromagnetic interaction. However, since Te-\textit{p} states form a nearly closed shell configuration, the interaction is very weak. Although the exact values of the exchange constants $J_i$ depend on the Hubbard repulsion parameter $U$, we can evaluate their strength by comparing with the experimentally obtained Curie-Weiss temperature, as recently demonstrated for another frustrated lattice \cite{PhysRevLett.125.027203}. The Curie-Weiss temperature is given by $\theta_{CW}$ = $-\frac{S(S+1)}{3k_{B}}\sum_{i} Z_iJ_i$, where \textit{Z}$_{i}$ represents the coordination numbers of Mn atom. By plotting $\theta_{ CW}$ as a function of \textit{U} (see Fig.~\ref{plotthe}) we find a match with the experimental value of $-$152 K for \textit{U} = 5.75 eV for which  $J_{1}$ = 4.6 K, $J_{2}$ = 4.2 K and $J_{3}$ =  $-$ 0.2 K  with  \textit{J}$_{1}$:$J_{2}$:$J_{3}$ = 1 : 0.92 : $-$0.04. These values are slightly larger than those ($J_{1}$ = $J_{2}$ = 3.1 K and $J_{3}$ = $-$ 0.6 K) found from the fit of spin waves detected by inelastic neutron scattering \cite{PhysRevB.102.094413}. The latter values however underestimate the Curie-Weiss temperature by yielding  a bit smaller values of exchange couplings.  

 \section{Discussion} A conventional 3\textit{D} antiferromagnet exhibits a symmetry breaking phase transition at a temperature close to the Curie-Weiss temperature, but  the titled compound  BMTO shows long-range order only at $T_{N}$ =  20 K despite much larger  $\theta_{CW}$ $\sim$ $-$ 152 K. This suggests that  BMTO is a moderately frustrated antiferromagnet with frustration index \textit{f} = 7.
 In this system, two structural reasons could explain the observed  long-range order: (i) Magnetic ions are in the high-spin state (\textit{S} = 5/2) and quantum fluctuations are  less  pronounced even though the Mn$^{2+}$ ions are arranged in 2\textit{D}  triangular  plane; (ii) The intra-plane and the inter-plane nearest-neighbor distances are the same which allows for sizeable  inter-plane interactions that forces the system to undergo a  long-range magnetic ordering. The origin of the  magnetic ordering in this strongly correlated system is examined within the framework of DFT.  Our calculations of exchange interaction reveal dominant antiferromagnetic  intra-layer exchange coupling $J_{ 1}$ = 4.6 K and a comparable inter-layer $J_{2}$ = 0.92 $J_{1}$. Furthermore, a very weak ferromagnetic inter-layer interaction exists with third nearest neighbor ($J_{3}$ = $-$ 0.04 $J_{1}$) due to double-exchange interaction via the linear path Mn-O-Te-O-Mn. 
  The magnetic specific heat data below the AFM transition  are well reproduced with Eq.(\ref{eqn:ga})
  suggesting the presence of magnon excitations. A broad maximum at 10 K in magnetic specific heat data suggests the presence of gapped magnon excitation in the ground state. Similar types of broad maxima were also observed in BiMnVO$_{5}$ and MnWO$_{4}$  which indicates Mn$^{2+}$ ions are subjected to  anisotropic magnetic interactions \cite{Chowki_2016,Hardy_2016}. In BMTO, the estimated magnon excitation gap is 1.4 K, a similar value of magnon excitation  gap is also observed in MnWO$_{4}$ \cite{PhysRevB.83.140401}.  The missing of entropy which is estimated as $\sim$ 18\text{\%} of the expected entropy for \textit{S} = 5/2 system is due to the presence of spin frustration and significant short-range spin correlations between  Mn$^{2+}$ spins already above $T_{N}$, which are however underestimated by our crude modeling of the lattice contribution to the specific heat. In fact, the evolution of $\mu$SR spectra in weak transverse field show that short-range ordering effects become apparent already  at 35 K $\simeq$ 2$T_{N}$, and gradually increase as the temperature approaches  to $T_{N}$. The ratio  of the $\mu$SR amplitudes $A_{0}$(\textit{T})/$A_{0}$(50 K) $<$ 1 in the temperature range 20 K $\leq$\textit{ T} $\leq$ 35 K suggests the presence of short-range spin correlation above the antiferromagnetic transition temperature.  If there are no  significant short-range spin correlations above the transition temperature, volume fraction of the sample does not change above $T_{N}$ as observed in weak transverse field $\mu$SR spectra of LiCrO$_{2}$ \cite{Ikedo_2010}. The short-range spin correlations reflect the presence of moderate spin frustration in the magnetic lattice of BMTO.  The zero-field $\mu$SR spectra show that below 35 K the spin lattice relaxation rate gradually increases, which is commonly observed in the vicinity of magnetic phase transition temperature. At \textit{T} $>$ 35 K, the temperature independent initial asymmetry can be associated with the paramagnetic nature of Mn$^{2+}$ spin.  The position of sharp maximum in the muon relaxation rate and the reduction of initial asymmetry (\textit{A}$_{0}$) to $A_{0}$/3 both occurs at \textit{T} = 20 K, which confirms a phase transition at this temperatures.  Finally, static internal fields are directly observed through the oscillations of muon asymmetry below $T_{N}$ and the evolution of the order parameter is reflected in the temperature dependence of these fields. 
 \vspace{0.3cm}   
 \section{Conclusion}
 The double perovskite BMTO crystallizes in the trigonal crystal symmetry R$\bar{3}$m,  wherein Mn$^{2+}$ ions form  two dimensional triangular layers with sizeable inter-layer exchange coupling. Our comprehensive results, well supported by first principle calculations reveal the presence of antiferromagnetic  long-range magnetic order below $T_{N}$ = 20 K. Below $T_{N}$, magnetic specific heat data suggest the presence of magnon excitations with a gap of approximately  1.4 K, which indicates the presence of magnetic anisotropy as commonly observed in classical Heisenberg systems.  Our zero-field and weak transverse field $\mu$SR results provide a concrete evidence of static internal fields in the  long-range ordered state below $T_{N}$ and short-range spin correlations above $T_{N}$.
 This frustrated triangular lattice antiferromagnet is also potentially  interesting  to uncover exotic ground state associated with quenched disorder in triangular lattice  by substitution of less electronegative cations at the tellurium site. Further studies on single crystals are required to shed more insight into the low energy excitations of this double perovskite based frustrated magnet.
 \vspace{0.5 cm}
 \section{ACKNOWLEDGEMENTS}
 We thank DST, India for the  PPMS facility at IIT Madras. The financial support from Science and Engineering Research Board, and Department of Science and Technology, India through research grants  is acknowledged. AZ acknowledges the funding by the Slovenian Research Agency through the Project No. N1-0148, J1-2461, and Program No. P1-0125. HL acknowledges the financial support by SNF through the Project grant 200021L-192109. 
 
  \bibliographystyle{apsrev4-1}
 \bibliography{BMTO}

\end{document}